# All-dielectric active photonics driven by bound states in the continuum

Song Han[1, 2], Longqing Cong[1, 2], Yogesh Kumar Srivastava[1, 2], Bo Qiang[2, 3], Mikhail V. Rybin[4, 5], Wen Xiang Lim[1, 2], Qijie Wang[2, 3], Yuri S. Kivshar[5, 6], and Ranjan Singh[1, 2*]

[1]*Division of Physics and Applied Physics, School of Physical and Mathematical Sciences, Nanyang Technological University, Singapore 637371, Singapore*

[2]*Centre for Disruptive Photonic Technologies, The Photonics Institute, Nanyang Technological University, 50 Nanyang Avenue, Singapore 639798*

[3]*Centre for OptoElectronics and Biophotonics, School of Electrical and Electronic Engineering & The Photonics Institute, Nanyang Technological University, Singapore, 639798, Singapore*

[4]*Ioffe Institute, St Petersburg 194021, Russia.*

[5]*ITMO University, St Petersburg 197101, Russia*

[6]*Nonlinear Physics Center, Australian National University, Canberra ACT 2601, Australia*

[*]E-mail: ranjans@ntu.edu.sg

Date: March 6th, 2018

**Abstract**

Recently emerged dielectric resonators and metasurfaces offer a low-loss platform for efficient manipulation of electromagnetic waves from microwave to visible. Such flat meta-optics can focus electromagnetic waves, generate structured beams and vortices, enhance local fields for sensing as well as provide additional functionalities for advanced MRI machinery. Recent advances are associated with exotic optical modes called bound states in the continuum, which can give rise to extremely large quality factors and supercavity lasing. Here, we experimentally demonstrate subwavelength active supercavities with extremely high-Q resonances that could be reconfigured at an ultrafast time scale. We reveal that such supercavities enable all-optical switching and modulation of extremely sharp resonances, and thus could have numerous applications in lasing, mode multiplexing, and biosensing.

In the last few years, all-dielectric metamaterials based on resonances were proposed as an alternative to the lossy metal-based photonic structures [1-23]. Although their building blocks are larger relative to the metallic counterparts, the dielectric resonators do not have Ohmic losses. Metamaterials are driven by strong resonances and their quality factor is an important parameter that determines the strength of light-matter interaction. The structures with high quality (Q) factors offer a novel way for strong localization of electromagnetic energy in near fields that allow effective sensors and other optical devices[9-12,22]. In extreme case, the Q factor diverges in so-called bound states in the continuum (BIC) recently studied experimentally[24-26]. Here we report extremely high-Q factors in dielectric metasurfaces driven by the BIC conditions and realize active all-dielectric supercavity modes that could be dynamically switched at an ultrafast timescale[27].

For the metadevice fabrication, we use tunable materials to set-up their functionalities and exploiting the optical pulses to tune the material properties[28]. Terahertz (THz) metasurfaces made of intrinsic semiconductors offer a unique ability for modulation of its refractive index upon excitation by short optical pulses. The optical pump with energy higher than the bandgap can introduce photo-doping, i.e. free charge carriers into the semiconductors[27]. Optical pump–THz probe studies in a variety of semiconductors revealed free-carrier-induced changes in the complex conductivity described by simple Drude model[29-31]. As a result, the photoconductivity of semiconductors reveals an ideal approach for the dynamic control of high-Q resonances by tuning the external pumping power and the photon energy. This would open a huge perspective for designing active metamaterial and photonic devices towards filling the THz gap.

We consider a design of periodic dielectric metasurface that supports BIC modes. Recently high-index dielectric cuboids were shown to possess a rich family of Mie-type resonances[32]. The building block of our metasurfaces is a silicon cuboid with width $W$, length $L$ and height $H$. The cuboids are arranged on a fused quartz substrate at the nodes of a square lattice with the lattice constant $P$ (Fig. 1). Such metasurfaces can be relatively easily manufactured by a wide range of fabrication techniques and size scales because of its stripe air-groove structure.

Several types of BIC states might arise in planar periodic structures. The first one is the symmetry-protected BIC at the Γ point, which turns to a leaky mode with even a weak symmetry mismatch[33]. Its spectral position and Q factor depend strongly on the incident wave angle. Recently a strong laser action was reported due to a resonance-trapped BIC, which is supported by the metasurface consisting of dielectric cylinder[34]. The resonance-trapped BICs are less sensitive to the light direction and its Q factor remains very high even when the structural parameters are detuned from the optimal conditions.

We calculate the transmission spectra dependence by tuning *L* while other parameters were kept constant. In the present study we will be focusing on the intensive low-frequency mode marked by blue dots in Fig. 2a. At certain parameters, the low-frequency mode possesses a vanishing nature which is a common attribute of BIC[24,26,33]. However, the mode manifests itself in the transmission spectra as a pronounced Fano resonance, which might transform transmission peak into a transmission dip[35]. In particular, the mode vanishing at the frequency *f* of about 0.57 THz and *L*=167 μm (Fig. 2a) is due to formation of double Fano transmission dips (see Supplementary Information). The detailed analysis shows that this region is not due to the BIC and is rather a supercavity mode[34]. In contrast, the mode vanishing at *f* of about 0.4 THz when the length tends to 300 μm has the divergent Q factor, which corresponds to the BIC.

We study the low-frequency BIC mode in details by means of rigorous coupled-wave analysis, which yields amplitudes of the periodic media eigenstates as a function of the frequency. In the metasurface with *L* of about 300 μm and *W*=210 μm, the supercavity mode is explained by the destructive interference between a propagating mode and an evanescent solution. Also the analysis of eigenmode amplitudes makes it possible to evaluate Q factor directly (Fig. 2b), because in the vicinity of resonance the amplitudes are described by the Lorentz function, in contrast to the transmission spectra with Fano resonances.

The active control of metasurface in the THz range is always a tradeoff between the operation bandwidth and the field amplification for enhancement of device performance. The fast signal modulation in time domain has a wide spectrum in frequency domain. In contrast, the field amplification requires strongly localized modes having a narrow spectrum. Our metasurface allows us to design the structures with Q factor ranging from tens to over $10^4$. In this study we choose Q of about several hundred that enable ultrafast devices.

Transmission spectra of metasurface are shown by black solid curve in Fig. 3a. The spectra demonstrate strong scattering with a narrow transmission window at 0.42 THz owing to the supercavity mode. A fluctuation of structure element parameters both in length and width shifts the frequency of the mode but does not affect essentially its Q factor (Figs. 3b, 3c). Additionally we study the variation of the incident angle revealing traces of the symmetry-protected BICs. We find that the variation of the incident angle does not change dramatically the Q factor of supercavity mode.

Our samples are made of high-resistivity silicon being widely used in THz optics due to its low loss and low dispersion at these frequencies. The designed all-dielectric metamaterials are silicon resonator arrays shown schematically in Figure 1. The transmission spectrum of the samples was obtained using a terahertz time-domain spectroscopy (THz-TDS) system. In Figs. 3d,e, the length of the resonators is manipulated (260 μm and 270 μm) by keeping the width constant at 210 μm. The extracted $Q$ factors of the two samples are 80 (length = 260 μm) and 72 (length = 270 μm), respectively. In Figures 3d and 3f, the width of the resonators is changed (180 μm and 210 μm) while the length remains fixed at 270 μm. The experimentally extracted $Q$ factors of the samples are 72 (width = 210 μm) and 115 (width = 180 μm), respectively. It is clearly observed that both the resonant frequency and the $Q$ factor can be tuned by tailoring the geometric size of the silicon resonator. Compared with the numerical simulations, the relatively weak transmission intensity and broadening of linewidth in the experiments originate from the limited scan length and resolution of the THz-TDS (see Supplementary Information). However the measured transmission spectra demonstrate peaks, which can be explained as leaky modes related to the symmetry-protected BICs.

As one of the most promising applications, active modulation of a metadevice with on-demand optical properties are required for applications in wave-front engineering, near-field control of electromagnetic radiation, and multi-channel optical data processing[1-3]. In our device, the silicon cuboids behave as the active material, which generates free carriers when pumped by 800 nm femtosecond laser beam. The optically induced free carriers enable the change of conductivity on the surface of silicon resonators with a penetration depth of about 5 μm[29, 31]. We experimentally obtain an all-optical active switching of the high-$Q$ supercavity mode where the switch is controlled by applying the external stimulus. As shown in Figure 4a, the supercavity mode is observed in the metasurface without optical pump, and the mode completely vanishes when pumped with fluence of 525 μJ/cm$^2$. The mode switching behavior is also clearly visible from time-domain signals in terms of the contrast in the long-lasting oscillations of the time domain THz pulse. Due to the existence of the high-$Q$ mode in the silicon resonator without pump, the THz photons remain confined inside the resonator (Figures 2c, 2d) that leads to a strong ringing (oscillations) in the terahertz pulse at longer time delay as revealed by the 210-picosecond high-resolution long scan time-domain signal (see inset in Figure 4a). Upon photo-excitation, the change of the conductivity on the silicon surface impacts the interaction of the incoming THz wave with the supercavity. Therefore, the ringing in the THz pulse decays rapidly to zero with an external photoexcitation and thus the supercavity mode disappears. As a verification of the modulation, we performed high-resolution numerical simulations and plotted the dynamic evolution by changing the surface conductivity of silicon resonators within a depth of 5 μm (Figure 4b). The high-$Q$ resonance in the metasurface is switched off for an extremely small value of photoconductivity (1000 S/m), indicating the highly sensitive nature of the supercavity mode.

In addition to the excellent switching performance of all-dielectric metasurfaces, we also probed the modulation of the resonance in the time-domain signal. The gradual damping is explicitly observed from the time-domain profiles by varying the pump fluencies (Figure 4c). The transmitted pulses under different pump fluencies demonstrate the functionality of gradual modulation in the proposed dielectric supercavity. As a dynamic switch, the operation speed is one of the most important parameters to characterize the performance. In our case, the switching speed is dominated by the relaxation time of photo carriers as per the dynamics shown in Figure 4d. At different pump fluencies, the rise time of photo carrier accumulation is about 10 picoseconds, however, the relaxation takes several nanoseconds so that we estimate the operation speed of the device to be in the gigahertz range. This switching speed could be improved by ion-implanted silicon, as well as utilizing direct band-gap semiconductors (GaAs), and/or intra-band effects in semiconductors as building blocks[27,36].

In summary, we have experimentally demonstrated an active control of high-*Q* terahertz all-dielectric supercavities with strong field confinement. We also observed BIC that enhanced the *Q* factor of supercavity mode by an order of magnitude to several thousands. The dynamic application of the device was also experimentally demonstrated by performing an all-optical active switching of supercavities. Our findings open an avenue for dynamic high-*Q* all-dielectric metadevices as well as passive functional metadevices operating as modulators, filters, and biosensors in the terahertz regime. The simple design can also be scaled down to nanoscale for optical applications, such as low threshold nano-lasers, on-chip parametric amplifiers, and harmonic generators.

**METHODS**

**Numerical simulations.** Numerical studies are carried out by using commercial finite-element frequency-domain solver COMSOL Multiphysics. The periodic boundary conditions are applied along the *x/y*-directions. A perfect matched layer is applied at the input and output ports.

**Sample fabrication**. The larger size of designed resonators operating in the THz band requires a different fabrication from those operating in the visible and IR spectrum. In particular, such a thick silicon wafer on insulator (SOI) could not be grown by chemical vapor deposition (CVD) and thus, it would be virtually impossible to fabricate these resonators by a conventional one-step reactive ion etching (RIE). Therefore, we followed three main steps to fabricate the samples: (1) patterning of photoresist spin-coated silicon wafer with UV lithography; (2) sticking the silicon wafer to a quartz substrate with UV curable adhesive (NOA 85, *n* = 1.46); (3) deep etching of silicon-on-quartz composite with RIE process [5]. This direct fabrication route for achieving a thick SOI structure is simpler than other all-dielectric metamaterials that needs to be embedded in a homogeneous low-index medium. 250 nm $SiO_2$ is deposited on the high-resistivity (>10000 Ω· cm) silicon wafer with thickness of 200 μm by the PECVD (plasma-enhanced chemical vapor deposition), where the selectivity of the etching gas ions on Si and $SiO_2$ determines the thickness of the SiO$_2$ layer. The hybrid silicon dioxide-on-silicon wafer is patterned by conventional mask photolithography on SiO$_2$ with a 500-nm layer of S1805 photoresist. The hybrid silicon dioxide-on-silicon wafer with patterned photoresist is then directly bonded on a 1-mm quartz substrate with a thin layer of spin-coated UV curable polymer optical adhesive (Norland Optical Adhesive 85) and exposure to 10 min of UV light (2.5 W/cm$^2$). The SiO$_2$ layer is then removed by mixed gases of CHF$_3$ and CF$_4$ and the remaining pattern is kept as a protective mask for subsequent etching. The silicon wafer is etched by the DRIE (deep reactive-ion etching technique; Oxford Estrelas). Each cycle of the Bosch process consists of sidewall passivation (C$_4$F$_8$) and etching (SF6) steps. Each cycle of the Bosch process consists of 5 s of deposition and 15 s of etching. In the deposition step, the C$_4$F$_8$ gas (85 sccm) is utilized with 600 W ICP power at 35 mTorr pressure. During the etching step, a mixture of SFhe etching step, a mixture of S6 (130 sccm) and O$_2$ (13 sccm) is applied with 600 W ICP power and 30 W bias power, at 35 mTorr pressure. This process cycle is then repeated until the silicon is completely removed. Solid silicon micro-cube arrays are kept attached to the fused silica substrate. The periodicities in *x* and *y* directions are 300 μm.

**Terahertz time-domain spectroscopy (THz-TDS) measurements.** Passive transmissions of metamaterial samples were characterized by a typical 8*f* confocal THz-TDS system with an operation bandwidth of 0.2 to 2.5 THz consisting of a pair of photoconductive antenna transmitter and receiver. The metallic antenna (transmitter) on GaAs chip is DC biased at 70 V, and excited by a near-infrared laser beam (Ti: sapphire oscillator laser system, 100 fs, 6 nJ per pulse at 800 nm with 80 MHz repetition rate). During measurements, the time domain transmitted THz signals through the samples were recorded with a scan length of 210 ps. The time domain signals were transformed to frequency domain and normalized by using an identical bare silicon substrate as the reference by $\left|\tilde{t}(\omega)\right|=\left|\tilde{E}_S(\omega)/\tilde{E}_R(\omega)\right|$, where $\tilde{E}_S(\omega)$ and $\tilde{E}_R(\omega)$ are the Fourier transformed spectra of sample and reference, respectively. The measurements were done at room temperature and in the dry nitrogen atmosphere to nullify the effect due to water vapor absorption.

**Optical pump-THz probe (OPTP) spectroscopy.** Dynamic performance of the all-dielectric metamaterials was performed using the optical pump-THz probe (OPTP) spectroscopy setup. A pulsed laser beam from an amplifier (120 fs, repetition rate 1 kHz, and beam diameter 6 mm) was coherently split into three parts: external pump source, pump beam for generation of terahertz radiation and detection beam for terahertz detection. For dynamic terahertz spectra measurements, the pump beam was delayed by 16 ps to capture the maximal modulation of the resonance mode, and the pump power was tuned by a neutral density filter. For the measurements of free carrier dynamics, we set the delay stage of terahertz probe at point II in Figure 4(c), and monitored the time-varied terahertz amplitude at this point by changing the arrival time of pump pulse on the sample.

**Acknowledgments**

The authors S.H., L.Q.C., Y.K.S., W.X.L., and R.S. acknowledge Singapore Ministry of Education (MOE), Grants MOE2011-T3-1-005, and MOE2015-T2-2-103. M.V.R. and Y.S.K. acknowledge a support by the Ministry of Education and Science of the Russian Federation (3.1500.2017/4.6) and the Australian Research Council.

**Competing financial interests**

The authors declare no competing financial interests.

**FIGURE CAPTIONS**

**Figure 1. Design of all-dielectric metasurface supporting supercavity modes.** (**a**) Schematic view of the all-dielectric metamaterial. The properties of the dielectric resonators are actively modulated by applying external optical pump. (**b**, **c**) Top and perspective view of the fabricated sample by scanning electron microscope, respectively. The geometrical parameters are: lattice constant $P$ = 300 µm, length $L$ = 270 µm (along $y$ axis), width $W$ = 210 µm (along $x$ axis), and height $H$ = 200 µm.

**Figure 2. Optimization of geometrical parameters.** (**a**) Parameter sweeping of the cuboid length from 60 µm to 300 µm, where the width (210 µm) and the height (200 µm) are kept constant. The low frequency mode is marked by blue dashed curve. (**b**) Q factor of the low frequency mode of the metasurface with $W$=210 µm as a dependence on the cuboid length. (**c**, **d**) Simulated electric and magnetic field profile showing tight field confinement inside the silicon metasurface, respectively. (**a**)-(**d**) The incident wave is polarized along the $x$ axis.

**Figure 3**. **Experimental studies of dielectric metasurface.** (**a**-**c**) Simulated normal transmission spectra of the supercavity mode are shown by black solid curves for the structure with parameters $L$=270 µm, $W$=210 µm (**a**), $L$=260 µm, $W$=210 µm (**b**), and $L$=270 µm, $W$=180 µm (**c**). Simulated transmission spectra averaged over incident angle in the range of -5° to 5° are shown by red dashed curves. Spectral positions of the symmetry-protected BICs are shown by grey dash verticals. (**d**-**f**) Experimental transmitted intensities of supercavity with parameters $L$=270 µm, $W$=210 µm (**d**), $L$=260 µm, $W$=210 µm (**e**), and $L$=270 µm, $W$=180 µm (**f**). The incident wave is polarized along $x$ axis.

**Figure 4. Active modulation of a supercavity device.** (**a**) Experimentally measured transmission spectra without and with optical pump to switch On and Off the high-$Q$ resonance mode, respectively. Inset: Transmitted THz time-domain pulses with and without optical pump. To clearly observe the long-lasting oscillations due to the excitation of the high-$Q$ resonance mode, the amplitude of time-domain signals is amplified by 20 times after 165 ps in the 2$^{nd}$ segment. (**b**) Corresponding simulated transmission spectra by changing the conductivity on the surface of silicon. (**c**) Measured transmitted THz time-domain pulse with varying pump fluencies. (**d**) Resonance recovery time studied by probing the transient amplitude change at point II (green arrow in Figure 4**c**). The normalized recovery dynamics was captured at point II by sweeping the pump pulse (@800 nm, 1.55 eV, silicon band gap ~1.14 eV) with time delay relative to the terahertz pulse.

**Figure 1**

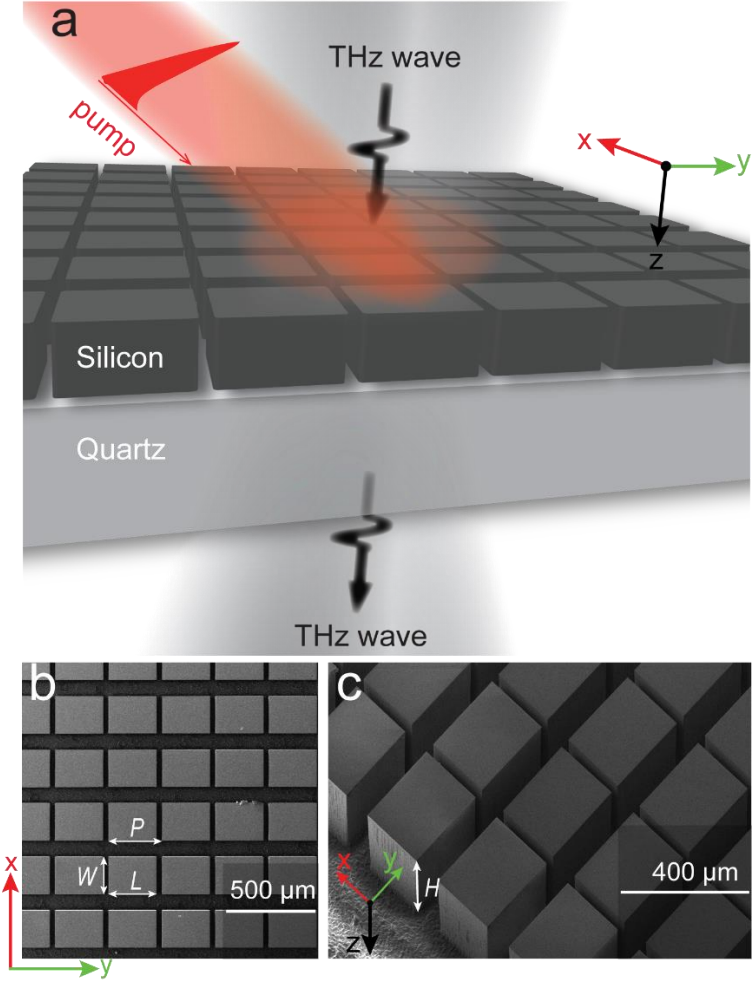

**Figure 2**

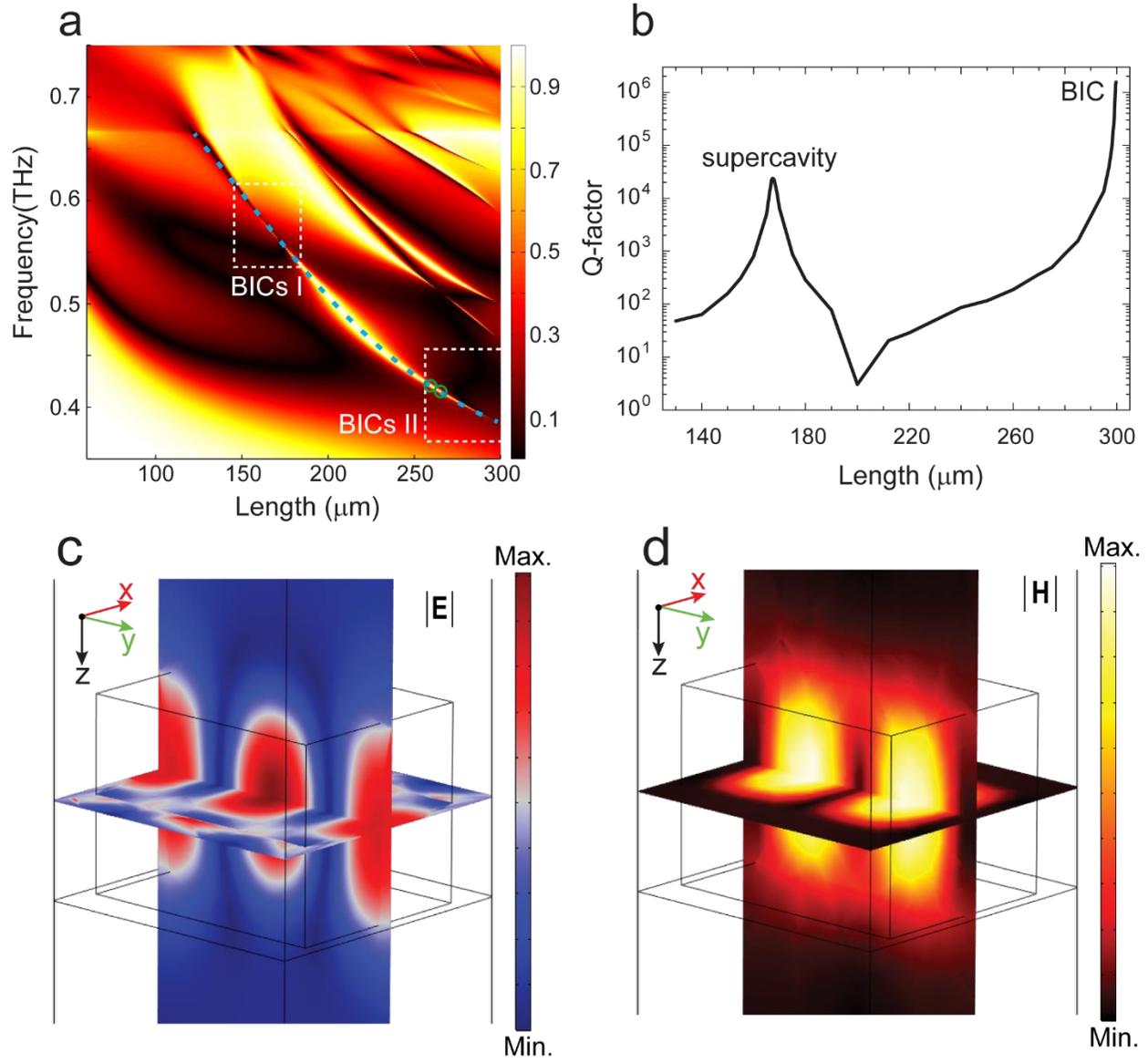

**Figure 3**

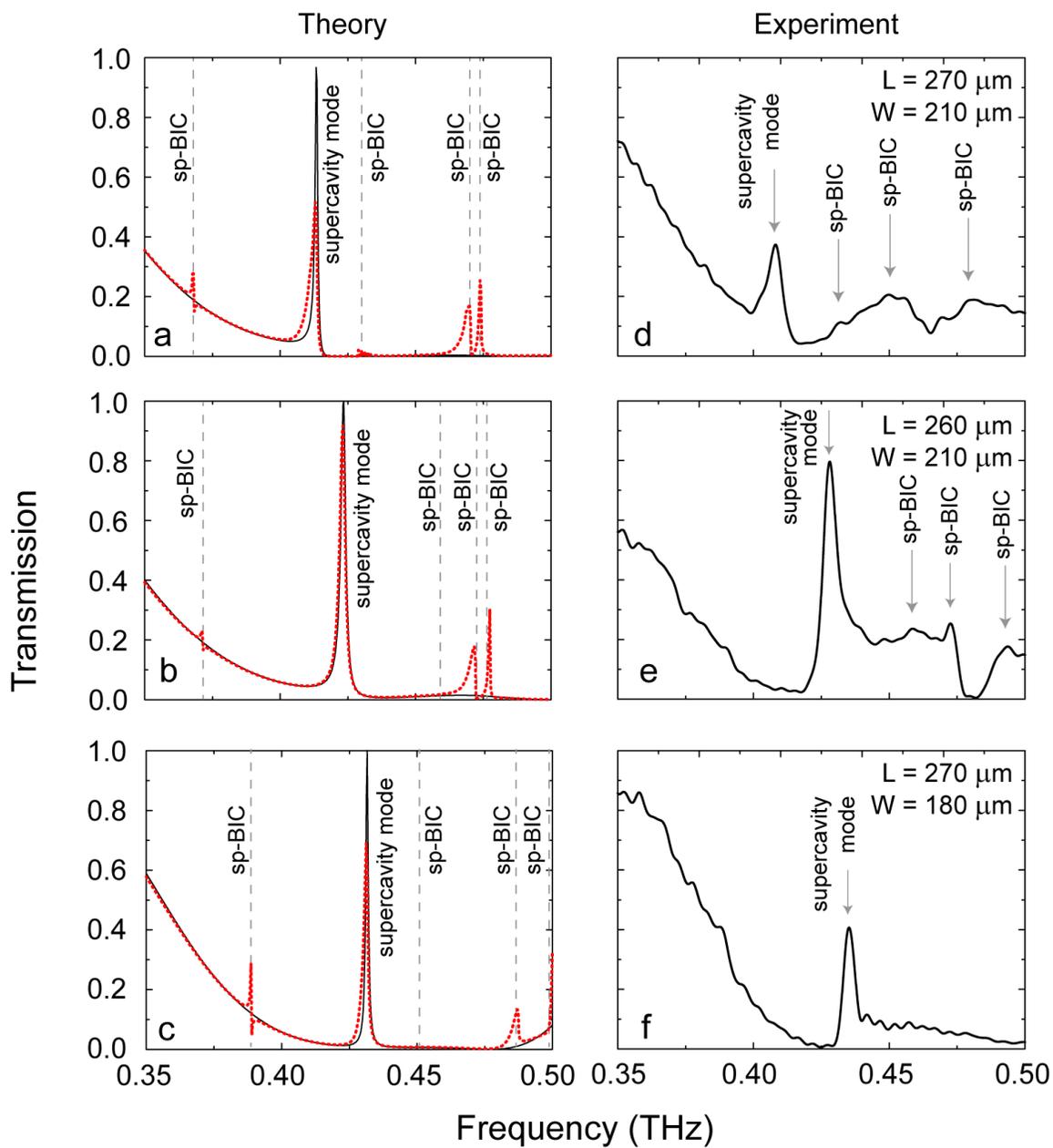

**Figure 4**

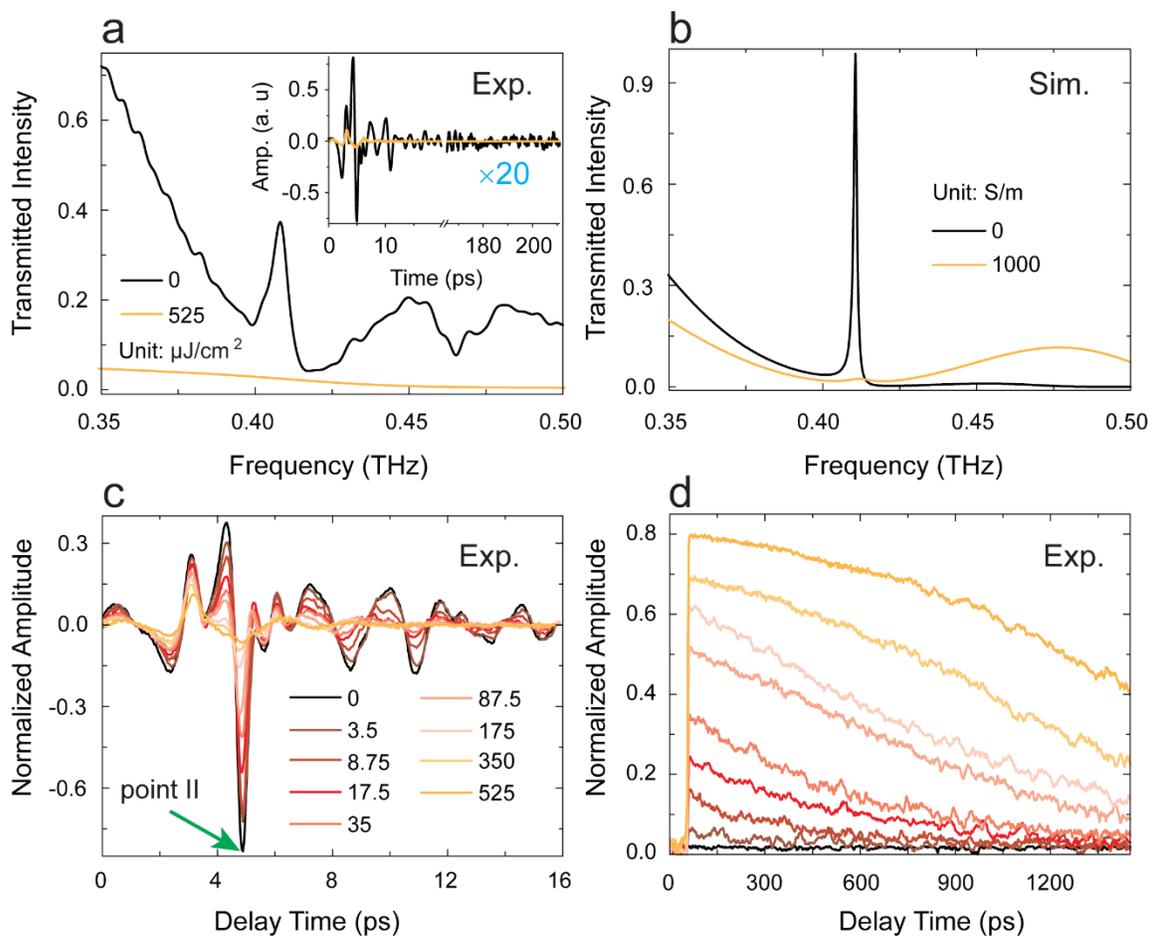